\documentclass[11pt]{article}

\usepackage{geometry}

\usepackage{amsmath,amsfonts,amsthm,amscd,amssymb}  % These are to write mathematics
\usepackage{dsfont}			% Símbolos para os conjuntos R, N, etc.
\usepackage{thmtools}		% Teoremas

\declaretheoremstyle[
spaceabove=6pt, spacebelow=6pt,
headfont=\normalfont\bfseries,
notefont=\normalfont\bfseries, 
notebraces={}{},
bodyfont=\normalfont\itshape
]{Estilo1}

\declaretheorem[style=Estilo1,numbered=no,name=\!\!]{ThmName}

\newcommand\ket[1]			{\left| #1 \right\rangle}
\newcommand\braket[2] 		{\left\langle #1 \mid #2 \right\rangle}
\newcommand\norm[1]			{\left|\left| #1 \right|\right|}

\newcommand{\Schrodinger}			{Schr\"o\-din\-ger}

\begin{document}

\title{Everettian Decoherent Histories and Causal Histories}

\author{Andr\'e L. G. Mandolesi \\ \emph{Departamento de Matem\'atica, Universidade Federal da Bahia} \\ \emph{Salvador-BA, Brazil} \\ \emph{E-mail:} \texttt{andre.mandolesi@ufba.br}}

\date{\today.}

\maketitle

\abstract{D. Wallace has tried to use decoherence to solve the preferred basis problem of Everettian Quantum Mechanics, and this solution lays the foundation for his proof of the Born rule. But this is a circular argument, as approximations used in decoherence usually rely on the probabilistic interpretation of the Hilbert space norm. 

He claims the norm can  measure approximations even without probabilities, but this assumption has not been properly justified. Without it, the combination of the Everettian and decoherent histories formalisms leads to strange consequences, such as a proliferation of small amplitude histories with lots of macroscopic quantum jumps.

Still, this erratic behavior may provide a way to justify the approximations, in a new histories formalism, in which macroscopic causal relations play a central role. Small histories, suffering too much interference, may lose causality, being thus discarded as invalid. The remaining branches can present some small interference, opening the possibility of experimental verification.
}

\section{Introduction}

Everettian Quantum Mechanics (EQM), or Many Worlds Interpretation \cite{DeWitt1973,EverettIII1957}, is an attempt to solve the measurement problem of Quantum Mechanics by rejecting the measurement postulate. It applies the rest of the usual formalism to all systems (even macroscopic ones), at all times (even during measurements). In EQM, a measurement is just a quantum entanglement of the measured system with the measuring device and the observer, which evolve into a macroscopic superposition of different versions of themselves, each registering one of the results. The collapse of the wavefunction is illusory, due to the fact that, as evolution is linear and interference is negligible, each version of the observer is unaware of the other components of the superposition (which are called \emph{branches} or \emph{worlds}).

EQM faces a \emph{preferred basis problem}: how to decompose a macroscopic quantum state into branches behaving approximately as the classical reality we observe. 
There is also a \emph{probability problem}: evolution via \Schrodinger's equation is deterministic, and all measurement results are obtained (even if in different branches), so why do quantum experiments seem probabilistic, following the Born rule (which was eliminated with the measurement postulate)?

An idea to solve this last problem, proposed by D.\,Deutsch \cite{Deutsch1999} and developed by D.\,Wallace \cite{Wallace2010,Wallace2012}, uses decision theory to show that rational agents, following EQM, should decide on bets about quantum experiments as if results were probabilistic and followed the Born rule. Such high level approach requires branches where agents exist as macroscopic quantum systems, so it depends on a solution to the first problem. Wallace \cite{Wallace2012} tries to solve it via decoherent histories, but approximations used in decoherence, on the basis of negligible probabilities, are invalidated without the Born rule. So these solutions form a circular argument \cite{Baker2007}.

Wallace claims the Hilbert space norm is a dynamically natural feature of the formalism, and  this justifies the approximations. 
But, as we show, he has not made this idea precise, nor provided a well developed defense of it. And, without a good justification for the approximations, his mix of decoherent histories and EQM might not deliver the expected results. 

Still, if EQM is ever to work, it must incorporate decoherence somehow. So we analyze what really results from \emph{Everettian Decoherent Histories (EDH)}: a combination of these formalisms, prior to the Born rule, and without assuming from the start the validity of the usual approximations. The consequences turn out to be quite strange. For example, without a reason to neglect small amplitude histories, we must consider the possibility that all histories do happen and are equally relevant, no matter how bizarre. Tiny ones may exhibit frequent macroscopic quantum jumps, leading to a breakdown of causality at the macroscopic level. On the other hand, all macroscopic states exist at nearly all times, in one history or another, suggesting a complementary perspective, in which nothing really changes at the macroscopic level. 

These results suggest EDH might not work, but they may actually carry the seed of a solution: tiny histories behaving so erratically as to lose causality should not be considered valid histories, in the sense of narratives in which events are meaningfully connected. 
So we propose the development of a \emph{causal histories} formalism, exploring relations between Born weights, interference and causality to get a more restrict definition of history. 
If successful, it might justify the approximations and solve the preferred basis problem, possibly paving the way for a solution to the probability problem.

Our results might also help settle the \emph{testability problem}: whether EQM can be distinguished experimentally from the usual Quantum Mechanics. If the causal histories formalism works, causal branches should be free from the worst results of EDH, but might exhibit tiny deviations from the usual predictions. This could, in principle, provide a way to test EQM, if our experimental capabilities ever get precise enough. If the formalism does not work, and those results turn out to be actual consequences of EQM, this theory should be discarded for disagreeing with observations. 

Section \ref{sec:Preliminaries} reviews the measurement problem, EQM and its problems, and the histories formalisms.
In section \ref{sec:Approximations and continuity} we criticize Wallace's arguments for using decoherence without probabilities, and in section \ref{sec:Everettian Decoherent Histories} we analyze its combination with EQM without the usual approximations. 
Section \ref{sec:Histories and macroscopic causality} questions the definition of history borrowed from the decoherent histories formalism, and proposes a more restrict one, stressing the importance of causal relations.
Section \ref{sec:Conclusion} summarizes our conclusions.

\section{Preliminaries}\label{sec:Preliminaries}

\subsection{The Measurement Problem}\label{sec:Measurement Problem}

In the Copenhagen interpretation of Quantum Mechanics (CQM), the measurement postulate states that, if a quantum system is in a state 
\begin{equation}\label{eq:psi}
\ket{\psi}=\sum_i c_i\ket{i},
\end{equation}
with $\braket{i}{j} =\delta_{ij}$ and $\sum |c_i|^2=1$, and is measured with respect to the basis $\{\ket{i}\}$, the result is one (and only one) of the $i$'s, and the state collapses to the corresponding $\ket{i}$. Also, results are probabilistic, according to the following rule.

\begin{ThmName}[Born Rule]
The probability of result $i$ is $p_i=w_i$, where $w_i$ is its \emph{Born weight}, 
\begin{equation} \label{eq:Born_weight}
w_i=|c_i|^2=|\braket{i}{\psi}|^2.
\end{equation}
\end{ThmName}

This postulate agrees with experimental data, but is
conceptually ambiguous. It sets measurements apart from other quantum processes, which obey the deterministic linear \Schrodinger\ equation, but lacks a precise definition of what are measurements. 
These might be distinguished for involving a classical macroscopic system, such as an observer, but if this system's particles obey \Schrodinger's equation, how can they collectively produce a nonlinear probabilistic process? Probabilistic even in principle, not simply due to lack of knowledge about the states of the particles. And how does the collapse of the quantum state happen? 
Many attempts have been made to solve this \emph{measurement problem}, such as hidden variables theories, Bohmian mechanics, nonlinear \Schrodinger\ equations, and others, each with its own difficulties \cite{Auletta2000,Wheeler2014}.

This relates to the problem of whether Quantum Mechanics remains valid as systems get bigger, with Classical Mechanics emerging from it. 
In the usual view, quantum superpositions should not happen at the macroscopic level, lest we observe \Schrodinger\ cats. But nothing in the quantum formalism seems to induce their disappearance in large systems, quite to the contrary. So many physicists consider Quantum Mechanics valid only for microscopic systems, with a new theory being needed to explain the quantum-classical transition. This view becomes problematic as quantum phenomena are verified at increasingly larger scales, or for research in fields like quantum cosmology.
Some see decoherence as an explanation for the emergence of classicality, but it is questionable whether it eliminates superpositions, or merely wipes out interference between their components, which remain nonetheless.

\subsection{Everettian Quantum Mechanics}\label{sec:EQM}

A solution, proposed by H. Everett III \cite{DeWitt1973,EverettIII1957}, rejects the measurement postulate, and applies the rest of the quantum formalism even to  macroscopic systems. Evolution is always deterministic, following \Schrodinger's equation, even during measurements.
It leads to macroscopic superpositions, but also explains why observers do not perceive them. If not for some unsolved problems, it might explain quantum measurements, and provide the link between quantum and classical mechanics. 

In EQM, measurements are just quantum entanglements of measuring devices with whatever is being measured. More precisely, a \emph{measuring device} for a basis $\{\ket{i}\}$ of a system is any apparatus, in a quantum state $\ket{D}$, which interacts in such a way that, if the system is in state $\ket{i}$, the composite state evolves as\footnote{For simplicity, we assume the system remains in state $\ket{i}$, but this is not necessary.}
\begin{equation}\label{eq:measuring device}
\ket{i}\otimes\ket{D}\ \  \longmapsto\ \  \ket{i}\otimes\ket{D_i},
\end{equation}
where $\ket{D_i}$ is a new state of the device, registering result $i$.
Linearity of Schr\"odinger's equation implies that, if the system is in state \eqref{eq:psi}, the composite state evolves as
\begin{equation*}
\ket{\psi}\otimes\ket{D} = \left(\sum_i c_i\ket{i}\right)\otimes\ket{D} \ \  \longmapsto\ \   \sum_i c_i\ket{i}\otimes\ket{D_i}.
\end{equation*}
This final state is to be accepted as an actual quantum superposition of macroscopic states. But it will not be perceived as such by an observer looking at the device, as, by the same argument, his state $\ket{O}$ will evolve into a superposition, according to
\begin{equation}\label{eq:MacroscopicSuperposition}
\left(\sum_i c_i\ket{i}\otimes\ket{D_i}\right)\otimes\ket{O} \ \  \longmapsto\ \   \sum_i c_i\ket{i}\otimes\ket{D_i}\otimes\ket{O_i},
\end{equation}
with $\ket{O_i}$ representing a state in which he saw result $i$. 
By linearity, each component $\ket{i}\otimes\ket{D_i}\otimes\ket{O_i}$ evolves independently, as if the others did not exist, as long as interference is negligible. This condition is usually justified, for macroscopic systems, using decoherence arguments.

Everett's interpretation of this final state is that the observer has split into different versions of himself, each seeing a distinct result. Each version evolves as if the initial state had been $\ket{i}\otimes\ket{D}\otimes\ket{O}$, so he does not feel the splitting, nor the existence of his other versions.
Each component is called a \emph{world} or a \emph{branch}, and this evolution of one world into a superposition of many is called  \emph{branching}. So in EQM all possible results of a measurement actually happen, but in different worlds. The observer in state $\ket{O_i}$ only thinks the system has collapsed into $\ket{i}$ because he does not see the whole picture, with all other results and versions of himself.

Problems that plague the Copenhagen Interpretation disappear in EQM, but new ones come along, as discussed below.

\subsubsection{Probability Problem}\label{sec:Probability_Problem}

In EQM, when measuring \eqref{eq:psi}, any result $i$ with $c_i \neq 0$ is obtained with certainty, even if not all versions of the observer see it. The \emph{probability problem} is reconciling this with experiments, which indicate results are probabilistic and follow the Born rule. 

It has a qualitative aspect, of how probabilities can emerge from a deterministic theory. In classical mechanics, processes can appear random due to our ignorance of details, but in EQM one must explain the apparent randomness even if the quantum state and its evolution are perfectly known.

There is also the quantitative aspect of accounting for probability values.
By a \emph{Born-like rule} we mean any result  explaining why, in an \emph{Everettian universe} (i.e. one governed by EQM), quantum experiments would appear probabilistic, with probabilities given by the Born weights \eqref{eq:Born_weight}. 
Many attempts have been made to obtain such result \cite{Albert1988,Buniy2006,EverettIII1957,Graham1973,Hanson2003,Zurek2005}. 

Deutsch \cite{Deutsch1999} proposed an adaptation of Decision Theory \cite{Karni2014,Parmigiani.2009} to EQM, to show that, in an Everettian universe, it would be rational to make decisions, related to bets on the results of quantum experiments, as if the outcomes were probabilistic, with Born weights playing the role of probabilities. 
The idea was further developed by Wallace, which presented a formal proof \cite{Wallace2010,Wallace2012}.
But use of decision theory requires worlds where narratives with agents, measurements and payoffs make sense. This involves solving first the next problem.

\subsubsection{Preferred Basis Problem}\label{sec:Preferred_Basis_Problem}

Decomposition of states like those in \eqref{eq:MacroscopicSuperposition} may not be unique \cite{Zurek1981}, so it is not clear whether there is one basis which gives the correct description in terms of worlds, or how to find it. 
Also, Everett's description of measurements assumes that EQM gives rise, at least in some branches, to macroscopic structures, like devices and observers, that behave classically, most of the time, up to a good approximation. 

The \emph{preferred basis problem} consists in finding a natural way to decompose the quantum state of a macroscopic system into branches which behave like the classical reality we observe (even if not all of them, and not all the time).
Wallace \cite{Wallace2012} has proposed an adaptation of the decoherent histories formalism to solve this problem. But, as we discuss in section \ref{sec:Approximations and continuity}, the probability problem may invalidate it.

\subsubsection{Testability Problem}\label{sec:Testability problem}

Solving those problems would put EQM in a better theoretical standing than CQM, and might shed some light on the \emph{testability problem}: whether experiments can tell these theories apart.

Many physicists disregard EQM for predicting the same observable results as CQM, and therefore not being testable. But if that is the case, any quantum experiment is a test of EQM as much as of CQM. Impossibility of testing which one better describes our universe does not make one worse than the other. Had EQM been developed first, CQM might be the one being disregarded for making no new predictions (besides being theoretically ambiguous). 

Granted, one may feel uneasy with a theory predicting undetectable other worlds. But any theory, even classical mechanics, has elements which can not be directly observed, but are accepted due to other consequences being confirmed.
 
Anyway, there is no proof that EQM and CQM are experimentally equivalent. Everett's description of measurements assumes unitary quantum mechanics gives rise to devices and observers, branches do not interfere, and the Born rule holds somehow. Validity of these assumptions depends on solving the previous problems, and answers we have so far indicate EQM might not precisely replicate CQM's predictions. 

In a worst case scenario, Wallace's proposed solution to the preferred basis problem might actually lead to predictions at odds with observations, as shown in section \ref{sec:Everettian Decoherent Histories}, indicating that EQM might simply be wrong. In the best case, perhaps such bad predictions can be dismissed via a solution to the probability problem or the formalism of section \ref{sec:Causal Histories}, and EQM can be tested through small deviations from the Born rule or some new macroscopic quantum phenomena. There have been proposals of how to do so \cite{Deutsch1986,Page1999,Plaga1997}, but they are beyond our present experimental capability.

\subsection{Histories Formalisms}\label{sec:Histories formalisms}

Wallace's approach \cite{Wallace2012} to the preferred basis problem is based on the decoherent histories formalism, which we review below. 

\subsubsection{Decoherence}\label{sec:Decoherence}

\emph{Decoherence} \cite{Joos2003,Schlosshauer2007,Zurek2002} is a process by which an open quantum system loses some quantum characteristics, as it interacts and gets entangled with its environment. 

Models show some states (called \emph{pointer states}) are more robust with respect to such interaction, i.e. have a stronger tendency to remain disentangled. We say such states are \emph{selected} by the environment, which measures them in the sense of \eqref{eq:measuring device}.  In some cases they form an orthogonal basis, but in others they constitute an overcomplete set of vectors, as for example in the quantum brownian motion, where they are minimum-uncertainty Gaussian packets (coherent states) $\ket{q,p}$. This example is considered a good paradigm for a macroscopic system, whose position and momentum are continuously measured (with some imprecision) through the scattering of particles from the environment. This would explain why macroscopic systems are observed in states of fairly well defined position and momentum. 

As the environment is differently affected by distinct pointer states, and such differences spread across its many degrees of freedom, it rapidly evolves into (almost) orthogonal states. Off-diagonal elements (\emph{coherences}) of the reduced density matrix of the system, with respect to the pointer states, decay extremely fast. This (almost) eliminates interference between such states, and we say the system has \emph{decohered}. 
Close pointer states take longer to decohere, but this problem can be reduced by coarse graining, as we discuss in section \ref{sec:Decoherent Histories}. 
As systems get bigger, it is hard to shield them from the environment, and decoherence becomes ever present.

As the reduced density matrix becomes (nearly) diagonal, it formally resembles a classical probabilistic mixture. This is considered an important step in the quantum-classical transition, but should not be interpreted as if the system had lost its quantum nature and become classical.
Tracing out the environment hid information about how it is entangled to the system, but the composite system remains in a pure quantum state. 
Adepts of CQM must explain the disappearance of all, but one, of the components of the mixture. 

In EQM, all components remain, but entangled to (almost) orthogonal states of the environment. 
In this view, decoherence is just  entanglement at work, a measurement of pointer states by the environment, which selects a basis for decomposition into branches, and (almost) eliminates interference between them. What distinguishes it is that it happens continuously, and, for all pratical purposes, is irreversible, as information spreads through the environment's many degrees of freedom.

The reduced density matrix of an open system follows a master equation, and its dynamics lacks unitarity, which is central to Wallace's proof of the Born rule. So in describing the evolution of branches he turns to the consistent/decoherent histories formalisms, which apply to closed quantum systems.

\subsubsection{Consistent Histories}\label{sec:Consistent Histories}

The \emph{Consistent Histories (CH)} formalism, conceived by Griffiths \cite{Griffiths1984,Griffiths2002} and further developed by Omn\`es \cite{Omnes1988,Omnes1999}, aims to describe quantum processes in ways which allow the use of classical (Boolean) logic and classical probabilities. It identifies conditions allowing us to assign classical probabilities to sets of alternative histories, conceived as sequences of events or propositions about a system.

Its point of view is opposite to the Everettian one: quantum evolution is always stochastic, and quantum states are only tools for calculating probabilities in a set of possible evolutions, only one of which happens. Still, parts of the formalism adapt well to the Everettian setting, if properly reinterpreted. We present a simplified version\footnote{Reflecting its stochastic point of view, CH is usually presented using density operators. Sometimes it includes a final state, or allows quantum sample spaces to be branch dependent.}, assuming a normalized pure initial state $\psi_0$ at time $t_0$. 

A \emph{quantum sample space} is an orthogonal projective decomposition of Hilbert space, i.e. a family $\{P_\alpha\}$ of orthogonal projection operators with $\sum_{\alpha} P_{\alpha}=\mathds{1}$ and $P_{\alpha} P_{\beta}=\delta_{\alpha\beta}P_{\alpha}$.
It represents an exhaustive set of mutually exclusive events or propositions about the system. 

A \emph{history space} is a sequence $\{P_{\alpha_1}(t_1)\}, \ldots, \{P_{\alpha_n}(t_n)\}$ of quantum sample spaces, at times $t_0<t_1<\ldots<t_n$.
Here $ P_{\alpha}(t)=U(t,t_0)^{-1}P_{\alpha}U(t,t_0)$, as in the Heisenberg picture, and $U(t,t_0)$ is the unitary time evolution operator of \Schrodinger's equation. 

A \emph{history} $\alpha=(\alpha_1,\ldots,\alpha_k,\ldots,\alpha_n)$ is a sequence of events, specifying one $P_{\alpha_k}(t_k)$ from each sample space. In CH, only one history happens in  each history space. And even if $\alpha$ happens, we can not say that, at time $t_k$, the system is in a state in the image of $P_{\alpha_k}(t_k)$, for the ontology of CH is based on histories, not quantum states. States are seen as artifacts of the formalism, and a different history space might yield another ``true'' history, whose projector at $t_k$ could even be orthogonal to $P_{\alpha_k}(t_k)$.

To a history $\alpha$ we associate a \emph{branch state vector} $\psi_\alpha=P_{\alpha_n}(t_n) \cdots P_{\alpha_1}(t_1) \psi_0$. In CQM, with sample spaces representing projective measurements, this would be the final state (up to normalization and time translation) if the sequence $\alpha$ of results happens, which has probability $p_\alpha=\norm{\psi_\alpha}^2$. In EQM, it is one of the branches resulting from the evolution of $\psi_0$, and we are yet to make sense of probabilities. 
In CH, it has no such interpretations, being only a tool to obtain the probability of history $\alpha$ happening, which, under appropriate conditions (consistency, as described below), is postulated to be, once more, 
\begin{equation}\label{eq:probability_history}
p_\alpha = \norm{\psi_\alpha}^2 = \norm{P_{\alpha_n}(t_n) \cdots P_{\alpha_1}(t_1)\psi_0}^2.
\end{equation}

A history space $\{\bar{P}_{\bar{\alpha}_k}(t_k)\}$ is a \emph{coarse-graining} or \emph{coarsening} of $\{P_{\alpha_k}(t_k)\}$ if each $\bar{P}_{\bar{\alpha}_k}(t_k)$ is a sum of $P_{\alpha_k}(t_k)$'s. Conversely, $\{P_{\alpha_k}(t_k)\}$ is a \emph{fine-graining} or \emph{refinement} of $\bar{P}_{\bar{\alpha}_k}(t_k)$. 
This concept is included in the formalism by bundling history spaces into a \emph{history algebra}, specified by  taking, at each $t_k$, a \emph{quantum event algebra}, which is a Boolean algebra\footnote{With operations $P_1\wedge P_2=P_1P_2$, $P_1\vee P_2=P_1+P_2-P_1P_2$, and $\neg P=\mathds{1}-P$.} of orthogonal projectors.
The Boolean condition implies projectors commute, so their ranges are mutually orthogonal if the intersection is $\{0\}$. 

Given a coarsening $\{\bar{P}_{\bar{\alpha}_k}(t_k)\}$ of $\{P_{\alpha_k}(t_k)\}$, each coarser history $\bar{\alpha}$ can be seen as a family of finer ones: we write $\alpha\in\bar{\alpha}$ if, for each $k$, the range of $\bar{P}_{\bar{\alpha}_k}(t_k)$ contains that of $P_{\alpha_k}(t_k)$. Then
$  \psi_{\bar{\alpha}}=\sum_{\alpha\in\bar{\alpha}} \psi_\alpha. $
The formalism requires the $p_\alpha$'s to behave as classical probabilities, so they must be additive,
$ p_{\bar{\alpha}} = \sum_{\alpha\in\bar{\alpha}} p_\alpha$. 
Hence there must be no interference between histories, and we require\footnote{The condition $\text{Re}\langle \psi_\alpha | \psi_\beta \rangle=0$ suffices for additivity, but is problematic for composite systems \cite{Diosi2004}.} $\langle \psi_\alpha | \psi_\beta \rangle=0$  for $\alpha\neq \beta$. 
History spaces satisfying this condition are called \emph{consistent}\footnote{The terminology varies. Wallace uses consistency for additivity of probabilities, and decoherence for non-interference (which relates, but is not equivalent, to the concept from section \ref{sec:Decoherence}).} (relative to $\psi_0$), and are the only ones allowed in CH. 

A consistent history space is seen as a valid, or allowed, sequence of questions about the system, at different times. In such space, only one sequence of answers, or history $\alpha$, turns out to be true, with probability $p_\alpha$. CH dismisses many quantum paradoxes by noting they involve inconsistent history spaces, so the problem lies in us asking an invalid set of questions. Of course, it is debatable whether it really solves the paradoxes, or just forbids us asking inconvenient sets of questions.

Two consistent history spaces can be \emph{incompatible}, in the sense that they can not be combined into a single consistent one. Either one can be used to describe a quantum process, but not both at once. This is the source of much criticism. If in each space one history actually happens, are there many equally real but incompatible histories? Is it possible to impose some condition that all ``true'' histories be compatible in some sense, even if their history spaces are not? Can the formalism be supplemented with new conditions singling out one consistent set? 

A history space has a \emph{branching structure} (relative to $\psi_0$) if histories do not merge after diverging, i.e. if $\alpha$ and $\beta$ satisfy $\alpha_i\neq\beta_i$ and $\alpha_j=\beta_j$, for some $i<j$, one of them has zero probability. Consistency and branching are related as follows
 \cite{Griffiths1993,Wallace2012}.

\begin{ThmName}[Branching-Consistency Theorem]\label{Branching-Consistency Theorem}\footnote{Branching-Decoherence Theorem, in Wallace's terminology.}
Any history space with a branching structure is consistent, and the converse holds for some consistent refinement of it.
\end{ThmName}

\subsubsection{Decoherent Histories}\label{sec:Decoherent Histories}

Classicality is more than its probability laws, and CH admits histories that are far from classical. 
The \emph{Decoherent Histories (DH)} formalism, developed by Gell-Mann and Hartle \cite{Gell-Mann1990,Gell-Mann1993}, combines CH and decoherence to get special history spaces with a more classical behavior. 
It seeks not only (approximate) consistency\footnote{In their terminology, medium decoherence.}, but also \emph{quasi-classicality}, meaning histories approximately follow classical equations of motion, interrupted at times (as in quantum measurements) by some quantum behavior. 

In DH, the space of relevant variables of the subsystem of interest is partitioned into cells $\Sigma_\alpha$, large in comparison with the coherence lenght (below which decoherence is not effective), yet small enough for the required precision. This determines a quantum sample space given by operators $P_\alpha \otimes \mathds{1}_{\text{env}}$, where
\begin{equation}\label{eq:P_alpha}
P_\alpha=\int_{\Sigma_\alpha} \mathrm{d}{x} |{x}\rangle\langle{x}|, 
\end{equation}
and $\mathds{1}_{\text{env}}$ is the identity operator for the environment's Hilbert space (with the environment taken to include the irrelevant variables of the subsystem). 
A history $\alpha$ specifies, at a sequence of times $t_k$, in which $\Sigma_{\alpha_k}$ the subsystem is\footnote{If necessary, different families of cells can be used at each time.}. 
Coarsenings and refinements are obtained varying the cell size.

While in CH consistency is a precondition to admit a history space, in DH it arises via decoherence. For large enough cells, states in the range of different $P_\alpha$'s are distinct enough to quickly get entangled to (almost) orthogonal states of the environment\footnote{This fails near cell boundaries, but if cells are not too large, distinguishing adjacent ones is irrelevant.}. Orthogonality tends to subsist as the environment evolves, for its many degrees of freedom keep \emph{records} of the history. For example, particles scattered by the system in different directions, at distinct $\Sigma_\alpha$'s, tend to remain in (almost) orthogonal states, and affect other degrees of freedom in distinct ways. 
Erasing from the environment all traces of the history is impossible in practice. 
Of course, as different states of the environment evolve, they can spontaneously develop similar components, blurring their records, but their Born weights would be negligibly small.   

As records ensure (almost) consistency, (almost) additive probabilities can be assigned to histories, as in CH.
Some interference between histories subsists, but it is negligible if they are coarse enough. Deviations in the additivity of probabilities should be irrelevant, as long as they are too small to be detected experimentally.

By the Branching-Consistency Theorem, some refinement of this (almost) consistent history space will have an (approximate) branching structure.
But such refinement will not be in terms of projectors $P_{\tilde{\alpha}}\otimes \mathds{1}_{\text{env}}$ into smaller cells $\Sigma_{\tilde{\alpha}}$, as it will require projecting environmental states onto the different records.

Models show that, if histories are coarse enough, and the system has enough inertia to resist noise from the environment, histories with non-negligible probabilities will (approximately) follow classical equations of motion, with a stochastic force.

So a (almost) consistent history space seems to emerge naturally in DH, and its non-negligible histories are quasi-classical. 
However, we can still have incompatible descriptions of the evolution, as shown by Dowker and Kent \cite{Dowker1996}.

\section{Approximations and Discontinuities in Everettian Quantum Mechanics}\label{sec:Approximations and continuity}

Wallace \cite{Wallace2012} uses DH to solve the preferred basis problem. But as EQM is not stochastic, and all histories happen, probabilities \eqref{eq:probability_history} can no longer be postulated. As approximations used in decoherence rely on the negligibility of tiny probabilities, to justify them one should first solve the probability problem. But Wallace's solution of this other problem depends on branches provided by the first solution, forming a circular argument \cite{Baker2007}.

He claims use of decoherence is valid even without probabilities, as its approximations could be justified in other ways. As we show in section \ref{sec:Histories and macroscopic causality}, this may turn out to be true, but the reason is not so immediate.
To get there, we first need to dismiss his arguments for accepting approximations from the start, so we can have the strange results of section \ref{sec:Everettian Decoherent Histories}, which justify tightening the definition of history. 

We are used to the idea that, whenever states $\psi_1$ and $\psi_2$ are very close in Hilbert space, i.e. $\norm{\psi_1-\psi_2}$ is very small, we can approximate $\psi_1\cong \psi_2$ in calculations or experiments. But for this to be valid, whatever concrete physical meaning we attribute to, or extract from, an abstract Hilbert space element must be continuous with respect to that space's metric. In other words, we need to know for certain whether states that are close in the metric represent similar physical situations.

 In CQM, a quantum state carries information about probabilities of measurements, and the Born rule allows us to ignore small components, corresponding to unlikely results. So the approximation $\psi_1\cong \psi_2$ is valid because replacing one state by the other causes only small changes in probabilities. 

In EQM, until the probability problem is solved, we can not simply assume that tiny perturbations in a state can not alter much its physical meaning.
If $\psi_1$ and $\psi_2$ are quantum states of a macroscopic system, their components in the branch decomposition indicate which worlds are present. This ontology seems discontinuous: even if $\norm{\psi_1-\psi_2}$ is small, one state can have components not present in the other, so their physical interpretations can differ by whole worlds, invalidating $\psi_1\cong \psi_2$. 

For example, let $\ket{\psi_\epsilon}= \sqrt{1-\epsilon}\ket{0}+\sqrt{\epsilon}\ket{1}$, where $\ket{0}$ and $\ket{1}$ represent distinct branches. Without a Born-like rule or another justification, we can not neglect $\ket{1}$ nor consider it any less relevant than the $\ket{0}$ component, no matter how small $\epsilon>0$ is.  So the physical meaning of $\ket{\psi_\epsilon}$ may change drastically when $\epsilon$ goes from $0$ (a single branch) to nonzero (two equally relevant branches). 

We call this a \emph{branch discontinuity}: arbitrarily small perturbations in a state can create lots of completely different branches, none negligible (until further notice). One might object that no reasonable physical theory can have discontinuities like these, but the truth is we do not know EQM to be a good theory. Without a definitive explanation of how it connects with experiments, we must admit the possibility that it might lead to really bad predictions. Hence that objection is not valid. 

In \cite[p.253]{Wallace2012}, Wallace argues that, even without a probabilistic interpretation, the Hilbert space norm can still be used as a measure of approximations or perturbations. He does not provide a coherent defense of this, throwing just a few loose ideas in a short Socratic dialogue. His argument seems to go vaguely like this: (unspecified) dynamical features make the Hilbert space norm natural (whatever this means), and this naturalness (somehow) justifies its use to measure approximations. 

There is not much to do to contest such vague idea besides pointing out flaws in the few statements presented in its defense. Note that we are not saying the Hilbert space norm can not measure approximations, just that there is no reason to assume from the start that it can. It is up to Wallace to provide a good reason for relying on such assumption, and as we show he fails to do so. According to him \cite[p.253]{Wallace2012}:

\begin{itemize}
\setlength{\itemsep}{6pt}

\item  ``We can think of the significance of the Hilbert space metric as telling us when some emergent structure really is robustly present, and when it's just a `trick of the light' that goes away when we slightly perturb the microphysics''. 

\vspace{3pt}

But this is precisely the problem: if tiny perturbations can create new branches, branch decomposition is not a robust structure. Metrics do not make emergent structures robust by fiat, it is up to the formalism to show robustness with respect to the metric. If it can not, the idea that it gives rise to that structure may simply be wrong.

\item ``\ldots the Hilbert space norm is a perfectly objective feature of the physics, prior to any considerations of probability''. 

\vspace{3pt}

It is not clear what he means by objective, or why he considers the norm to be so.
In CQM, the norm is connected to probabilities, which can be considered objective in the sense that they can be measured. 
In EQM, it is not clear if the norm has observable consequences, so, instead of an objective feature, it might be just a mathematical artifact of a possibly flawed formalism.

\item ``What makes perturbations that are small in Hilbert-space norm `slight'\ldots? Lots of dynamical features of the theory. Small changes in the energy eigenvalues of the Hamiltonian, in particular, lead to small changes in quantum state after some period of evolution. Sufficiently small displacements of a wavepacket lead to small changes in quantum state too.''

\vspace{3pt}

Instead of clearly stating which relation between norm and dynamics is relevant for such `slightness', Wallace just gives examples where small changes in a physical situation cause small changes in the mathematical object representing it (quantum state). But tiny physical changes can also cause large changes in Hilbert space (e.g. a physically negligible displacement in a highly peaked wavepacket may send it to a nearly orthogonal state). More importantly, these examples do not imply that small changes in Hilbert space must represent small physical changes.  In fact, in EQM it seems tiny changes in quantum state might represent drastic physical changes (introduction of whole new worlds). If this seems unreasonable, then, again, it might just mean EQM does not work.

\item ``Ultimately, the Hilbert-space norm is just a natural measure of state perturbations in Hilbert space, and that naturalness follows from considerations of the microphysical dynamics, independent of higher-level issues of probability.'' 

\vspace{3pt}

He does not say which considerations are these, nor what being `natural' involves. Perhaps he means the Hilbert space norm is natural in the sense of being preserved by the dynamics, so small perturbations remain small as the system evolves.
But why would this be relevant for approximations? Classically, tiny changes in a system can grow as it evolves, yet nonconservation of the metric does not prevent its use to measure approximations.
In CQM, norm conservation is important for its probabilistic meaning, but states can grow increasingly dissimilar even as they stay at constant distance in Hilbert space (e.g. non-overlapping wavepackets moving apart). And in EQM it seems physically similar states might evolve into drastically different ones (with different worlds), even if their distance in Hilbert space remains small.

\item ``\ldots there's nothing Everett-specific about the problem \ldots I don't think there's any profound difference here between the role of the Hilbert space metric in quantum physics and, say, the spatial metric in classical physics.'' 

\vspace{3pt}

This may be a bad choice of example, as the spatial metric fails in both aspects that seem to interest him: it is not dynamically conserved, and states which are spatially close can be physically quite distinct (e.g. with different velocities).

Anyway, the situation in EQM is quite different from other theories. In CQM or classical mechanics, for example, a clear connection between theory and experiment tells us that close abstract states correspond to physically similar situations. We understand what states represent, how to measure their properties, and that these are continuous in the metric: close states give similar results. This is what makes some metric a good measure of approximations, not some vague `naturalness'. 
In EQM, however, the usual link between quantum formalism and observation has been severed. Until one manages to confirm the experimental meaning of a macroscopic state $\psi$ in EQM, there is no guarantee that close states, in the Hilbert space metric, would generate similar observations.  

\item ``Instantiation is always approximate, and we measure that approximation using the natural distance measures of the instantiating theory. There's need for self-consistency -- that distance measure had better appear natural from the perspective of the instantiated theory. But that self-consistency requirement doesn't make the whole enterprise viciously circular -- it can't, on pain of undermining science in general, not just Everettian quantum theory.''

\vspace{3pt}

Again, he throws around the term `natural' without saying what it means (that the distance is part of the theory?), or how it relates to approximations. 
This argument seems to repeat previous ones: the underlying theory has a distance measure, so whatever emerges from it must be continuous in that distance. But, again, we are not sure a branching structure emerges robustly from EQM, and, unlike with science in general, we do not know how EQM relates to observations.

\end{itemize}

In \cite[p.129]{Wallace2012}, Wallace mentions Everett's original proof of the Born rule \cite{EverettIII1957}, which was never accepted precisely because, without a probabilistic interpretation, there was no reason to neglect a low Born weight set of deviant branches (in which statistics deviate from the Born rule)\footnote{Graham's proof \cite{Graham1973} faced the same difficulty.}. And he suggests an argument to justify such negligibility: 
\begin{quote}
	\ldots the branching structure is [not] given exactly, by some particular choice of basis\ldots the structure is a consequence of decoherence, the branches are only approximately orthogonal, and indeed the theory only approximately specifies them. This suggests that in fact, at the emergent level appropriate to a branching description there may \emph{be} no deviant low-weight worlds. If their weight is small compared to the level of noise in the decoherence process, they will simply be an artefact caused by false precision, and will not have the robustness that\ldots was an essential requirement for emergent structure. As such (we might hope) \emph{all} of the branches will exhibit approximately correct statistics.
\end{quote}

Let us examine in which sense the emergent branching structure is approximate. A source of imprecision is the choice of cells: shifting their boundaries transfers Born weight between branches, and smaller cells can split a branch into several. For small enough cells, close ones are physically so similar that this imprecision is irrelevant.
A quite different approximation is the assumption that branch interference is negligible: it involves discarding tiny components in the evolution of branches. Wallace seems to mix the two, but the first imprecision does not justify this other approximation.
For example, as a branch evolves, its wavefunction can develop a tiny tail spreading across distant cells. Changing the cells gives a different decomposition of the tail into tiny branches, but does not eliminate it.

Wallace does not seem to consider his argument conclusive, as he then mentions Hanson's theory of mangled worlds \cite{Hanson2003,Hanson2006} as an interesting approach along those lines, but which needs more work\footnote{Our concept of causal histories (section \ref{sec:Causal Histories}) could also be included in this category.}.
Actually, he does not seem totally convinced by any of his arguments, conceding that ``\ldots there's plenty of work to be done here by philosophers of science with an interest on emergence'' \cite[p.254]{Wallace2012}. 
In fact, if he could really justify the negligibility of tiny components then Everett's proof would be valid, and he would not need to develop a more tortuous one via decision theory.
Unfortunately, his proof has the same problem as Everett's, and if that negligibility is not accepted for one then it can not validate the other either.

Still, even if he thinks this point requires more work, he does not seem to consider it a serious problem. After devoting only a few paragraphs to discuss it, throwing some loose ideas without properly developing them, he proceeds as if the question had been satisfactorily settled or had no significant impact on his proof of the Born rule. Unfortunately, as we show in section \ref{sec:Everettian Decoherent Histories}, the consequences of not being able to justify the usual approximations of decoherence can be quite severe.

\section{Everettian Decoherent Histories}\label{sec:Everettian Decoherent Histories} 

Without a good justification for its approximations in EQM, decoherence might not give its usual results. But entanglement with the environment still happens anyway, so it is important to analyze its consequences for EQM.

By \emph{Everettian Decoherent Histories (EDH)} we mean the parts of the DH formalism which do not depend on probabilities, reinterpreted from an Everettian perspective. In it, branch discontinuities come into play, usual approximations are no longer valid, and we must reanalyze all those ``approximate'' and ``almost'' permeating section \ref{sec:Decoherent Histories}. As \eqref{eq:probability_history} can no longer be postulated to be a probability, we call $\norm{\psi_\alpha}^2$ the Born weight of history $\alpha$. All histories with nonzero Born weight happen, and must be considered equally relevant (at least until there is some justifiable reason, like a Born-like rule, to treat these weights as measures of relevance).

The range of \eqref{eq:P_alpha} consists of wavefunctions with support in $\Sigma_\alpha$. As evolution via \Schrodinger's equation does not preserve compact supports, they instantly develop tails spreading across all cells. 
In CQM this is not seen as strange, for such tail gives a negligible probability that the system will jump to a distant cell. In EQM, it means new branches, corresponding to all other cells,  pop into existence. Without a Born-like rule, they must be considered as real and relevant as any. 

As a result, a multitude of highly non-classical histories, which were negligible in DH due to their tiny Born weights, become relevant. Essentially all histories do happen, some ill behaved even by quantum standards, with lots of macroscopic quantum jumps. Even from quasi-classical branches there is a continuous sprouting of tiny weird sub-histories.  The ``penalty'' for such bad behavior is a drastic reduction in Born weight, but, without a Born-like rule, this may be irrelevant.

One might argue that macroscopic quantum jumps can not happen, as they violate conservation laws. But these apply to averages over the whole state, not branchwise. Of course, if EQM does allow large violations in branches, and no Born-like rule is obtained to render them unlikely, then it should be discarded a failed theory. 

This compromises the very idea of measurement. A device designed to work as in \eqref{eq:measuring device} will always malfunction in some branches, spilling out all sorts of random results.
Again, this is not an acceptable description of our Universe, unless it is complemented by a Born-like rule allowing us to neglect most possibilities. But, until such rule is obtained, this is the picture we have to deal with in EDH.

It can be elucidating to compare this with Feynman's formalism of path integrals. One considers at first all possible paths, no matter how erratic. Destructive interference clears the picture, allowing us to consider only the more well behaved ones. But the more complicated ones are not really eliminated, they just have tiny Born weights. Without a Born-like rule, they can no longer be neglected.

Decoherence does not really happen in terms of cells with precise boundaries, as pointer states $\ket{\pi_x}$ tend to be like gaussian packets, peaked at a point $x$, but with small tails across all space. It does not make much difference in DH, as the tails give negligible probabilities, but in EDH it might be better to redefine \eqref{eq:P_alpha} as
\begin{equation}\label{eq:pointer projector}
P_\alpha=\int_{\Sigma_\alpha} \mathrm{d}{x} |{\pi_x}\rangle\langle{\pi_x}|.
\end{equation}
Evolution of the $\ket{\pi_x}$'s can be more stable, but now the ranges of the $P_\alpha$'s are only almost orthogonal to each other: a state in the range of $P_\alpha$ can have a tiny component in the range of another $P_{\alpha'}$. As it decoheres, besides $\alpha$ there will also be a tiny  (but equally relevant) component where the environment records $\alpha'$.

Evolution of the environment also comprises all possibilities, including tiny components where its macroscopic state changes drastically, and records disappear or get replaced by wrong ones. 
As records are not reliable, tiny branches can suffer significant interference from larger ones. For this, the difference in Born weights must be of a great many orders of magnitude, even more so if the branches and their histories are very different.

\subsection{Quasi-classicality and Dissimilarity}\label{sec:Dissimilarity}

In DH, non negligible histories behave quasi-classicaly. In EDH, no history is negligible, most exhibit strange behaviors, but some will be quasi-classical. 
However, even these might be nothing like the world we know.

Our macroscopic reality is built with atoms and molecules, whose stability and behavior depend on the Born rule, used in CQM to explain decay rates, chemical reactions, etc. 
Quasi-classicality is no substitute for this. 
A branch can approximately follow classical equations without even having stable atoms. For example, a branch where all atoms have disintegrated, and all particles behave, at the macroscopic level, like a classical gas, fits the definition of quasi-classical. 

Disintegration of all atoms, or other phenomena usually disregarded as unlikely, causes a huge reduction in  Born weight, so large branches can not have previously presented such problems. But, without a Born-like rule, preferring these would be akin to cherry picking branches similar enough to our world to give the desired results.

In EDH, the only reason to assume the existence of branches similar to our world is that everything will happen in at least some branches. But these will not be typical in any sense, not even among quasi-classical ones.

This presents a big obstacle for decision theoretic proofs of a Born-like rule, which depend on the existence of high level structures (agents, experiments, etc.), and on their behavior being somewhat similar to what we are used to. 
Any such proof should take into account the following condition:

\begin{ThmName}[Dissimilarity]\label{cd:Dissim}
Even quasi-classical branches can not be assumed to be similar to our macroscopic reality. 
So no assumptions can be justified, and no possibilities excluded, based on our physical experience. 
\end{ThmName}

For example, histories where a broken glass spontaneously becomes whole again should happen in EDH, and some may even be quasi-classical, so Dissimilary may encompass even the Second Law of Thermodynamics.

\subsection{Almost Orthogonality}\label{sec:Almost Orthogonality}

Orthogonality acquires its role in CQM via Measurement Postulate, which states that measurements\footnote{We refer only to projective measurements, as more general ones, like POVMs, can be reduced to them.} are in eigenvector bases of Hermitian operators, and uses orthogonal projections to define Born weights. Orthogonal states are mutually exclusive, in the sense that they can be eigenvectors for different values of an observable, and measuring one eigenvector never results in the other values. 
In CH, orthogonality is present in the projectors used to define histories, and in the requirement of consistency. In DH it appears in the projectors \eqref{eq:P_alpha}, which are an idealization of the more realistic, almost orthogonal, operators \eqref{eq:pointer projector}. Almost consistency is achieved by entanglement with almost orthogonal states of the environment.

In EQM, one must ask whether branches, which are emergent structures, turn out orthogonal to each other. EDH suggests we can have almost orthogonal measurements or branchings, since emergence of branches is more natural in terms of the almost orthogonal projectors \eqref{eq:pointer projector}, selected by decoherence with the environment. 
Perfect orthogonality can not be imposed just so EQM replicates conditions from CQM (like measurements in orthogonal bases), nor as a necessity to obtain a Born-like rule (as implicit in Wallace's Boolean condition \cite[p.175]{Wallace2012}\footnote{See \cite{Mandolesi2018} for a discussion.}). 

In questioning orthogonality, we must also ask why observables correspond to Hermitian operators, or why histories are defined with orthogonal projectors. 
Unitary evolution requires a Hermitian Hamiltonian, but it is not clear how this translates into branchings along eigenbases of Hermitian operators.
A measurement in EQM is a normal quantum process, which happens to lead to different versions of measuring device and observer, entangled to distinct states of the measured system, and evolving independently. Linearity of \Schrodinger's equation seems to allow this even for a nonorthogonal basis.

For example, let $\ket{0}$ and $\ket{1}$ be orthonormal states, and $\ket{+}=(\ket{0}+\ket{1})/\sqrt{2}$. It is not immediate why,  in EQM, no device $D$ can measure in the basis $\{\ket{+},\ket{1}\}$, i.e. interact in such a way that
\begin{equation}\label{eq:Nonorthogonal_Measurement}
\ket{+}\otimes\ket{D}\ \  \longmapsto\ \  \ket{+}\otimes\ket{D_+}, \quad \text{and}\quad
\ket{1}\otimes\ket{D}\ \  \longmapsto\ \  \ket{1}\otimes\ket{D_1}.
\end{equation}
By Dissimilarity, one can not argue no known device does this, or that our knowledge (based on CQM) forbids it. And, without  a Born-like rule, one can not say the probability for $\ket{+}$ would be greater than 1 in a measurement of $\ket{0}$ by $D$,
\begin{equation*}
\ket{0}\otimes\ket{D} = \left(\sqrt{2}\ket{+}-\ket{1}\right)\otimes\ket{D} \ \  \longmapsto\ \   \sqrt{2}\ket{+}\otimes\ket{D_+}-\ket{1}\otimes\ket{D_1}.
\end{equation*}

A valid objection is that, if $\ket{D_+}$ and $\ket{D_1}$ are macroscopically distinct, differing in the states of lots of particles, they must be almost orthogonal, so \eqref{eq:Nonorthogonal_Measurement} violates unitarity.
But this would not apply if, instead of $\{\ket{+},\ket{1}\}$,  $D$ measured some almost orthogonal basis  $\{\ket{a},\ket{b}\}$, provided \eqref{eq:measuring device} is adjusted to allow for measurements perturbing the basis states. We could very well have
\begin{equation*}
\ket{a}\otimes\ket{D}\ \  \longmapsto\ \  \ket{\tilde{a}}\otimes\ket{D_a}, \quad \text{and}\quad
\ket{b}\otimes\ket{D}\ \  \longmapsto\ \  |\tilde{b}\rangle\otimes\ket{D_b},
\end{equation*}
where $\ket{\tilde{a}}, |\tilde{b}\rangle$ are perturbed states satisfying
$\langle\tilde{a}\mid\tilde{b}\rangle\braket{D_a}{D_b}=\braket{a}{b}$.

This last condition requires $\ket{a}$ and $\ket{b}$ to be even closer to perfect orthogonality than the macroscopicaly distinct states $\ket{D_a}$ and $\ket{D_b}$. Can such tiny deviations from orthogonality be of any consequence?
Once we have a Born-like rule, the answer might be no, but until then, it is yes, for a couple of reasons: 
\begin{enumerate}
\item Mutual exclusivity is lost. If there can be a measuring device $D$ for an almost orthogonal basis $\{\ket{\epsilon},\ket{1}\}$, where $\ket{\epsilon}=\sqrt{1-\epsilon}\ket{0}+\sqrt{\epsilon}\ket{1}$ for small $\epsilon> 0$, then a measurement of $\ket{0}$ by $D$ will result $\ket{1}$ in some branches. 

\item Any set of mutually orthogonal states is linearly independent, but this fails with almost orthogonality, if the number of states is large. A linear combination of a huge number of almost orthogonal, macroscopically distinct, states, might even result in a state macroscopically quite different from all of them\footnote{One can build a large number of wavepackets peaked in a group of cells, whose sum peaks far away. The example is artificial, but shows the need to explain why it could not happen in realistic cases.}. So a quantum state might admit two decompositions in terms of quite different sets of worlds. Another way to look at this is that the set of branches which can emerge from EQM includes incompatible decoherent history spaces. 
\end{enumerate}

\section{Histories and Macroscopic Causality}\label{sec:Histories and macroscopic causality}

The problems of EDH may stem from the definition of history, which works well for DH, but becomes too broad when we no longer have probabilities to prune the worst branches. A solution might be to tighten it by requiring a sense of causality between events of a history. This may be quantified in terms of how little interference it suffers, with its Born weight providing an indication of its resistance to interference. An adequate formalism based on this idea is still in the works, so here we just lay out the main notions and possible ways to develop them.

\subsection{All Histories or No History?}\label{sec:All Histories or No History}

Even if its Born weight concentrates in some cells, a quantum state usually has components in all cells at nearly all times, as only exceptionally the wavefunction vanishes in a whole cell.  In DH, probabilities allow us to neglected most cells and focus on those of sizable Born weight. In a visual analogy, imagine cells in shades of gray, those of larger Born weight being darker. As dark spots move and divide into lighter ones, we get a picture of branching histories, each as probable as its final darkness level, with only quasi-classical ones being discernible.

In EDH, unless Born weights acquire some meaning, all nuance disappears, and cells are either black (nonzero component) or white (absolutely no component). Barring exceptional cases, they are black all the time. In this case, is it reasonable to say all histories happen, or there is no history? Sticking to a visual analogy, is a black paper full of black paths, or has it no paths?

Formally, we can define histories or paths as we wish. If paths are defined as arbitrary sequences of black dots, a black paper is full of paths. 
But we usually hope such labels reflect our intuitive ideas, which matter for interpreting results. For paths, we might want a color different from the paper, and maybe some notion of continuity to connect the dots. In transposing the definition of histories from DH to EDH, we must question whether it retains its intuitive interpretation. 

In the usual definition of a history $\alpha$, consecutive projectors $P_{\alpha_i}(t_i)$ and $ P_{\alpha_{i+1}}(t_{i+1})$ are linked by $U(t_{i+1},t_i)$, and a component in $\Sigma_{\alpha_{i+1}}$ only counts if it came from $\Sigma_{\alpha_i}$. 
In EDH, we must consider that usually a state in $\Sigma_{\alpha}$, at $t_i$, generates components in all $\Sigma_{\alpha'}$'s  at $t_{i+1}$, and, conversely, each $\Sigma_{\alpha'}$ will have components coming from all $\Sigma_{\alpha}$'s. Even if $U(t_{i+1},t_i)$ shows $\alpha_i$ contributes a component to $\alpha_{i+1}$, we can not say event $\alpha_{i+1}$ happens at $t_{i+1}$ because of $\alpha_i$ at $t_i$. Other cells would cause $\alpha_{i+1}$ anyway, and the component coming from $\alpha_i$ may even interfere destructively, reducing the Born weight of $\alpha_{i+1}$ (not that it matters, at this point).

So macroscopic causal relations get lost, if everything happens, all the time, in some branch, and interference prevents tracking, in a meaningful way, the causes and consequences of anything. 
Consider for example the following events:
\begin{itemize}
\item[A.] There is no glass anywhere;
\item[B.] All glasses are safely stored inside the kitchen cabinet;
\item[C.] There is a glass at the edge of the kitchen table;
\item[D.] There is glass shattered on the floor beside the kitchen table. 
\end{itemize}

Presented with D, one might think it most likely follows from C. But in EDH it may just as well result from B, with the glass breaking after tunneling from the cabinet, or even A, with atoms spontaneously joining to form pieces of glass on the floor.
In fact, two opposite points of view seem valid in EDH:
\begin{itemize}
\item All histories: there is glass on the floor for all these reasons at once, as any macroscopic state is generated by all others. 
\item No history: there is glass on the floor because, at all times, there is always glass on the floor in some branches.
\end{itemize}

The first one follows the definition of histories from DH. That it seems unreasonable is an indication that such definition may be too broad for EDH. The second one abandons that definition, and adopts the intuitive idea that histories should involve change, with events ceasing to happen and new ones coming about.

\subsection{Causal Histories}\label{sec:Causal Histories}

With a more adequate concept of history it may be possible to avoid such extremes, and also obtain a non-probabilistic interpretation for Born weights. 

Suppose, at time $t_1$, a state has its Born weight equally divided among the events A, B and C above, and, a little later, at $t_2$, the weight on D is not too small. As discussed, we can not say D happens at $t_2$ because of any specific event at $t_1$. But we can attribute its weight mostly to C, since A and B generate tiny components on D, which interfere negligibly with the larger one coming from C.

Imagine, on the other hand, that at $t_1$ most weight is on A or B, while C has a tiny component whose contribution to D is as small as that of A or B. Now interference keeps us from identifying any of these events as the source of the weight on D.

So Born weights play a non-probabilistic role, measuring resistance to interference, and allowing the establishment of causal relations for larger branches, less affected by interference. 
This brings shades of gray back to our visual analogy: wavepackets of large Born weight, suffering little interference, move along darker quasi-classical paths, amidst an ocean of almost white cells, whose lightness flickers, in indiscernible ways, as they suffer interference from everywhere. 
Such cells remain in a state of perpetual but timeless existence: they are present at (nearly) all times, but their Born weights, and their microstates, fluctuate meaninglessly, with no causal relations connecting them to other cells in a significant macroscopic narrative. A cell will only be part of an evolving narrative once a discernible wavepacket passes through it, linking it causally in time with other cells. 

This idea suggests a new formalism of \emph{causal histories}, including only those with Born weight large enough to resist interference and sustain macroscopic causal relations. Histories of low Born weight are discarded not on probability grounds, but for lack of causality: sequences of unrelated events are not histories in any meaningful sense of the word, only artifacts of a definition that was too broad. 

The Born weight below which histories are to be discarded can not be absolute. As they keep on branching, eventually even the largest one will fall below any fixed threshold. As its weight becomes much smaller than the total weight of the rest, one might think that interference inevitably becomes a problem. But interference decays rapidly with the distance, so cells that are far apart should not interfere significantly, unless their weight difference is of a great many orders of magnitude. So a tiny wavepacket can evolve causally as long as there are no large ones nearby.
Also, note that distances increase rapidly in high dimensional spaces: if, for example, $10^{30}$ particles are displaced by $10^{-10}$m each, the total system moves $100$\,km in configuration space. 

We want to say a history $\alpha=(\alpha_1,\ldots,\alpha_n)$ is \emph{causal} if any interference it suffers is so small that nearly all Born weight in cell $\Sigma_{\alpha_n}$, at time $t_n$, comes via $\alpha$.
Making this precise requires an appropriate measure of interference $I(\alpha)$, for which a history $\alpha$ will be causal if $I(\alpha)\ll 1$.
Note that a history can suffer significant interference at some steps, which latter dissipates, and a stronger causality condition might require negligible interference at each step.

We present some possible interference measures, which compare the effect of all other histories on $\Sigma_{\alpha_n}$ with that of $\alpha$: 
\begin{align*}
I_1(\alpha) &=\sum_{\substack{\tilde{\alpha}\neq\alpha, \\ \tilde{\alpha}_n=\alpha_n}} \frac{\norm{\psi_{\tilde{\alpha}}}}{\norm{\psi_\alpha}}, &
I_3(\alpha) &=\sum_{\substack{\tilde{\alpha}\neq\alpha, \\ \tilde{\alpha}_n=\alpha_n}} \frac{\left|\braket{\psi_{\tilde{\alpha}}}{\psi_\alpha}\right|}{\norm{\psi_\alpha}^2}, \\
 I_2(\alpha) &=\frac{\norm{P_{\alpha_n}(t_n)\psi_0-\psi_\alpha}}{\norm{\psi_\alpha}}, &
I_4(\alpha) &=\frac{|\braket{P_{\alpha_n}(t_n)\psi_0-\psi_\alpha}{\psi_\alpha}|}{\norm{\psi_\alpha}^2}.
\end{align*} 
Clearly, $I_3\leq I_1$ and $I_4 \leq I_2$. Also, $I_2\leq I_1$ and $I_4 \leq I_3$, as $P_{\alpha_n}(t_n)\psi_0=\sum_{\tilde{\alpha}_n=\alpha_n} \psi_{\tilde{\alpha}}$. So $I_1$ imposes the strongest condition, while $I_4$ gives the weakest one.
Still, if a history is $I_3$-causal (resp. $I_4$-causal), a refinement will be $I_1$-causal (resp. $I_2$-causal).

$I_1$ and $I_2$ take into account all components generated on $\Sigma_{\alpha_n}$, while $I_3$ and $I_4$ ignore those orthogonal to $\psi_\alpha$. This last approach may be preferable, since if $\tilde{\alpha}_n=\alpha_n$ but $\psi_\alpha$ and $\psi_{\tilde{\alpha}}$ are orthogonal they do not really interfere. For example, they might have different records in the environment, which (assuming they are correct) allow us to distinguish in  $\psi_\alpha+\psi_{\tilde{\alpha}}$ which part is caused by which history, and even refine them to become causal in the sense of $I_1$ and $I_2$.

$I_1$ and $I_3$ neglect interference between other histories, allowing lots of tiny $\psi_{\tilde{\alpha}}$'s, which might cancel out, to be counted as if they had a large cumulative effect on $\Sigma_{\alpha_n}$.  $I_2$ and $I_4$ take such interference into account, but this can also cause difficulties. Let $\tilde{\alpha}$ and $\hat{\alpha}$ be histories, distinct from $\alpha$, with $\psi_{\tilde{\alpha}}=\psi_\alpha$ and $\psi_{\hat{\alpha}}=-\psi_\alpha$. Should we say $\psi_{\tilde{\alpha}}$ and $\psi_{\hat{\alpha}}$ cancel out and the branch in $\Sigma_{\alpha_n}$ is caused by $\alpha$, or do $\psi_{\alpha}$ and $\psi_{\hat{\alpha}}$ cancel out and $\tilde{\alpha}$ causes $\Sigma_{\alpha_n}$? Are both $\alpha$ and $\tilde{\alpha}$ causal, or neither one?

With $I_1$, causal histories form an approximate branching structure, and with $I_3$ this holds for some refinement (possibly involving environmental variables). The same might be false with $I_2$ or $I_4$, as in the situation above.

Environmental records make $\langle \frac{\psi_{\tilde{\alpha}}}{\norm{\psi_{\tilde{\alpha}}}} \mid \frac{\psi_\alpha}{\norm{\psi_\alpha}} \rangle$ tend to $0$ quite fast, as $\tilde{\alpha}$ and $\alpha$ become more different. So, with $I_3$ or $I_4$, for $\alpha$ not to be causal there must be an $\tilde{\alpha}$, with $\tilde{\alpha}_n=\alpha_n$, for which $\norm{\psi_{\tilde{\alpha}}} \gg \norm{\psi_\alpha}$ by a great many orders of magnitude, all the more so if the histories are very different. So a history should be $I_3$- or $I_4$-causal whenever its Born weight is not too small, or if there are no close histories of much larger weight (nor more distinct ones with an extremely larger weight). Taking refinements if necessary, the same holds for $I_1$ or $I_2$.

An appropriate threshold for the inequality $I(\alpha)\ll 1$, above which histories are to be discarded as non causal, is also necessary. It should not be too arbitrary, and its precise value should not affect results significantly. This may be hard to come by, so we might have to consider instead a gradation of causality (which would bring the question of what does it mean for a history to be more or less causal).

If an adequate causality condition can be established, and the formalism works as expected, it will give Born weights an initial non-probabilistic role in EQM. This can be a step towards proving a Born-like rule, which, once obtained, might justify the choice of threshold, as one which allows probabilities of causal histories to fluctuate, due to interference, only within a desired precision. 

As causal histories can suffer tiny amounts of interference, this formalism opens the possibility of experimental verification for EQM. However, it should require a precision beyond our present capabilities.

Hanson \cite{Hanson2003} has also proposed using interference to justify discarding small branches, but through a different process: interference would cause small worlds to be ``mangled'', with observers ceasing to exist or remembering events from larger worlds. This seems like a quite dramatic interpretation of the concept of interference, and he provides no explanation of how such mangling process would happen.

\section{Conclusion}\label{sec:Conclusion}

Wallace's use of decoherent histories to solve the preferred basis problem is promising, but incomplete. Unless the probability problem is solved first, the Born rule can not justify the usual approximations in decoherence, and his attempt to justify them on non-probabilistic grounds is not satisfactory. 
Without a good reason to assume the validity of the approximations, one must admit that mixing EQM and decoherent histories might have weird consequences:
\begin{itemize}
\item Branch decomposition is not a robust feature of the formalism, due to branch discontinuities: arbitrarily small changes in a quantum state can create lots of completely different new branches. 

\item Nearly all macroscopic histories happen, and those of small Born weight are (until further notice) as relevant as large ones. 

\item All macroscopic states are present at nearly all times. So there is no real history, in the sense of a narrative with events coming into existence and ceasing to occur. 

\item Histories with tiny Born weights can exhibit frequent macroscopic quantum jumps, destroying any macroscopic sense of causality. 
Even those allowing a meaningful macroscopic narrative keep sprouting lots of ill behaved subhistories. 
 
 \item Quasi-classicality is not enough for similarity to our world. Tiny quasi-classical branches might not even have stable atoms.
 
\item Environmental records are unreliable, and tiny branches can suffer significant interference from much larger ones.

\item Measurements or branch decompositions in almost orthogonal bases are possible, and orthogonal states might not be mutually exclusive upon measurement. 

\item Even allowing for coarsenings or refinements, branch decompositions might be non unique, and two decompositions can involve quite different sets of branches.
\end{itemize}

Of course, if EQM really leads to such results, it must be discarded as a bad model for physical reality. But there may be ways to avoid this:
\begin{enumerate}
	\item Proving a Born-like rule would allow us to neglect small branches, which are the ill behaved ones, as unlikely. But if the proof depends on well behaved branches obtained through decoherence, the problems above must be considered, to avoid circularity. This may rule out decision theoretic proofs, and the consequences for Wallace's proof are analyzed in \cite{Mandolesi2018}.
	\item Developing a causal histories formalism, in which small branches that suffer so much interference and behave so erratically as to lose causality at the macroscopic level are not considered valid histories.
\end{enumerate}

Such formalism is under development, and many questions need answer to ensure it would work.
We must determine the most appropriate way to measure interference, and how much of it is too much for causality. Possibly there is not a unique answer, and distinct approaches might be better suited for different situations. 

If it works as expected, Born weights would acquire their first role in EQM, as measures of causality or resistance to interference. 
Also, causal histories turn out to be not only quasi-classical, but also to have all ingredients needed to form our classical reality: stable atoms, working molecules, etc. Basically, all phenomena deemed likely in CQM due to large Born weights should be present, while all the weirdness associated with small weights should be mostly absent. 

This solution of the preferred basis problem, and the fact that Born weights would finally have some significance in EQM, might provide an appropriate framework for proving a Born-like rule, via decision theory or some other approach. And, as causal histories are not completely free of interference, this opens, at least in principle, the possibility of testing EQM.

\bibliographystyle{amsalpha} 
\bibliography{../../Bibliografia_Many_Worlds/Bibliografia_Wallace}

\end{document}